\documentclass{article}

     \usepackage[final]{neurips_2019}

\usepackage[utf8]{inputenc} 
\usepackage[T1]{fontenc}    
\usepackage{hyperref}       
\usepackage{url}            
\usepackage{booktabs}       
\usepackage{amsfonts}       
\usepackage{nicefrac}       
\usepackage{microtype}      
\usepackage{array}
\usepackage[pdftex]{graphicx}

\title{Breaking Speech Recognizers to Imagine Lyrics}

\author{%
  Jon Gillick  \\
  University of California, Berkeley\\
  \texttt{jongillick@berkeley.edu} \\
\And
   David Bamman \\
   University of California, Berkeley\\
   \texttt{dbamman@berkeley.edu} \\
}

\begin{document}

\maketitle

\begin{abstract}
  We introduce a new method for generating text, and in particular song lyrics, based on the speech-like acoustic qualities of a given audio file.  We repurpose a vocal source separation algorithm and an acoustic model trained to recognize isolated speech, instead inputting instrumental music or environmental sounds.  Feeding the ``mistakes'' of the vocal separator into the recognizer, we obtain a transcription of words \emph{imagined} to be spoken in the input audio.  We describe the key components of our approach, present initial analysis, and discuss the potential of the method for machine-in-the-loop collaboration in creative applications.
\end{abstract}

\section{Introduction}

In songwriting and music composition, artists often develop their work through cycles of experimentation and feedback.  A composer may start by recording one instrument, later adding new parts or lyrics inspired by listening back to the first part.
As with many creative endeavors, feedback cycles like this sometimes result in writer's block.  A growing body of work on machine-in-the-loop frameworks [1, 2] seeks to assist artists in these situations by developing intelligent systems that surface possible variations or additions to the current state of the work.  To serve as a useful creative tool in practice, a computer system should make contributions that are related to the existing work along some dimension, but which still have enough diversity or randomness to be sufficiently different from the ideas already developed by the user [1].  Depending on the task, this can be a difficult balance to strike.  

In this work, we explore the case of lyric generation in songs. In practice, an important aspect of good lyric writing is the match between the words and the music, both sonically and semantically [3]; evidence suggests that even grammatically incorrect English lyrics written by non-native speakers have achieved great commercial success because of the way in which the \emph{sound} of the words fits with the music [4].
A system for lyric generation, then, should stand to benefit from some mechanism for taking a musical context into account.  While previous work on lyric generation has considered melodies as context [5], other connections between sound and lyric have yet to be explored computationally.  Drawing on the tradition of assisted orchestration in music [6, 7], we use machine learning to uncover latent \emph{lyrical} qualities inherent to sounds.  

\begin{figure}
    \begin{centering}
    \includegraphics[scale=0.385]{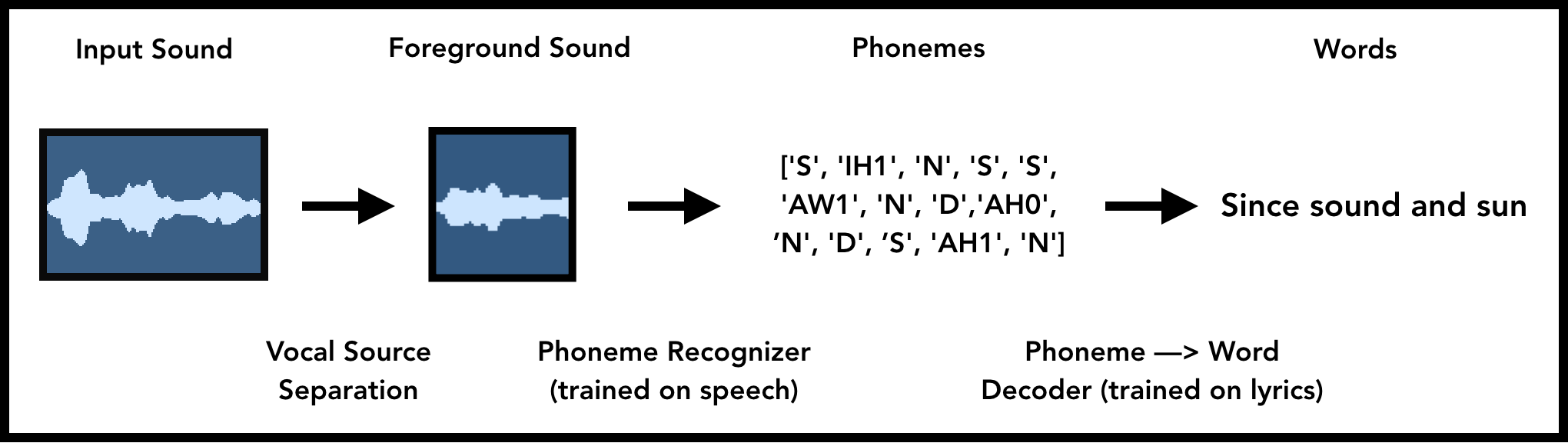}
    \label{fig:pipeline}
    \caption{Pipeline for imagining lyrics in music or environmental sounds.}
    \end{centering}
\end{figure}


\section{Method}

Our method consists of a three-stage pipeline.  Given an audio file, we apply, in order, \textbf{(1)} a vocal source separation algorithm, \textbf{(2)} a phoneme recognizer trained on clean speech, and \textbf{(3)} a language model trained on song lyrics.  Figure \ref{fig:pipeline} displays the series of transformations we use to generate text based on a sound. Additional implementation details can be found along with our code, which is available at \url{https://github.com/jrgillick/imagined-lyrics}.

\paragraph{Vocal Source Separation} We employ the REPET-SIM music/voice separation algorithm [8] using the implementation in the Librosa library [9].  This approach is based on the assumption that vocal sounds in recorded music appear in the \emph{foreground}, demonstrating different patterns of repetition over short time frames than background instrumentation.  When applied outside of its intended context, the algorithm tends to extract artifacts that, to our ears, have speech-like characteristics.

\paragraph{Phoneme Recognition} Most Automatic Speech Recognition (ASR) systems 
attempt 
to filter out any audio that is not speech; a good ASR system will not transcribe anything when given music or environmental sounds.  For our purposes, however, we are specifically interested in obtaining transcriptions for non-speech audio (which would be considered false positives in ASR). To this end, we train an RNN-based acoustic model to recognize phonemes in the LibriSpeech corpus [10]. Importantly, every training example we provide contains at least 5 phonemes, and we do not employ any additional methods to teach the model to discriminate between speech and non-speech sounds.

\paragraph{Phoneme to Word Decoder} The final step in the process is to decode the phonemes from the output of the acoustic model into words.  Here, we train a sequence-to-sequence model on lines and couplets from the Metrolyrics database,\footnote{\url{http://www.metrolyrics.com}} using the CMU pronunciation dictionary [11] to break words into phonemes.  During training, we randomly replace or drop out some of the ground truth phonemes in order to better approximate the noisy output of our acoustic model.  By replacing the text data in this model, the system could be adapted to target a different style of text.

\begin{table}[ht]
\centering
\caption{Examples of lyrics generated based on different sounds.}
\begin{tabular}[t]{l>{\raggedright}p{0.39\linewidth}>{\raggedright\arraybackslash}p{0.39\linewidth}}
\toprule
Input Sound&Lyric Samples&\\
\midrule
Shaker& ``Since sound and sun since innocence'' & ``So this sense thats sense standin'' \\
\hline
Bells&``Artist noticing artist testing''&``Exhibit in images is tearing''\\
\hline
Birds&``Happy the alley on one what will''&``Lil little wayne will live when i live'' \\
\hline
Music&``Tied wasted beside another dance we invented''&
``Healing howls understands relentlessly losers souls'' \\

\bottomrule
\end{tabular}
\end{table}%

\section{Discussion}

In initial experiments with a few different types of audio inputs, we find that \textbf{(1)} our method  produces very different outputs for different types of sounds, and \textbf{(2)}, subjectively speaking, the generated lyrics appear related to the timbre and prosody of the given sound.  For example, inputting a recording of a high-frequency shaker yields lyrics emphasizing the letter ``S''.  Often, the stressed syllables in the lyrics match up with the rhythm of the audio; this effect can be heard especially clearly when we apply our method to a recording of a musical instrument with speech-influenced phrasing, such as Peter Frampton's Talkbox-processed guitar.\footnote{Examples and an interactive demo can be found here: \url{http://bit.ly/imagining_lyrics_demo}}  These observations suggest that the methods introduced here hold the potential for creative use.  In future work, we hope to further expand the system to include an interactive interface in order to facilitate user studies.  

\section{Ethical Considerations}

Any text generation application that learns from data runs the risk of producing biased or offensive content reflective of the training data---our work, which learns a language model from song lyrics, is no exception.  One approach that we implement is to moderate the words shown to our language model based on a list of keywords\footnote{We filter out lyrics containing words from the list aggregated here: https://github.com/zacanger/profane-words}; a downside of this moderation is the potential for increased bias as a side effect of the content filter.  Recent work in modeling the intentions of vulgar expressions has shown promise for more nuanced profanity filtering or hate speech detection [12], and as these approaches improve, they present one path toward better automated moderation when working on text generation.  In addition to biased or offensive content, the other common ethical concern when working with material like lyrics is the potential for plagiarism.  In our setting, one of the reasons we choose to work with arbitrary audio inputs is to explicitly condition the language model with a strong signal which, intuitively, should make it difficult to copy from the training set; still, more work is needed to determine the extent to which the audio conditioning reduces plagiarism. 

\section{Acknowledgments}

The research reported in this article was supported by the Hellman Family Faculty Fund and by resources provided by NVIDIA. 

\section*{References}

\bibliography{ImaginingLyrics}

\small

[1] Clark, Elizabeth, Anne Spencer Ross, Chenhao Tan, Yangfeng Ji, and Noah A. Smith. ``Creative writing with a machine in the loop: Case studies on slogans and stories.'' In {\it 23rd International Conference on Intelligent User Interfaces}, pp. 329-340. ACM, 2018.

[2] Garfield, Monica J. ``Creativity support systems.'' In {\it Handbook on Decision Support Systems 2}, pp. 745-758. Springer, Berlin, Heidelberg, 2008.

[3] Pattison, Pat. {\it Writing better lyrics}. F+W Media, Inc., 2009.

[4] Seabrook, John. {\it The song machine: Inside the hit factory}. WW Norton \& Company, 2015.

[5] Watanabe, Kento, Yuichiroh Matsubayashi, Satoru Fukayama, Masataka Goto, Kentaro Inui, and Tomoyasu Nakano. ``A melody-conditioned lyrics language model.'' In {\it Proceedings of the 2018 Conference of the North American Chapter of the Association for Computational Linguistics: Human Language Technologies, Volume 1 (Long Papers)}, pp. 163-172. 2018.

[6] Cella, Carmine-Emanuele, and Philippe Esling.  ``Open-source modular toolbox for computer-aided orchestration.'' Timbre Conference. 2018.

[7] Gillick, Jon, Carmine-Emanuele Cella, and David Bamman. ``Estimating Unobserved Audio Features for Target-Based Orchestration.'' In {\it Proceedings of the 20th International Society for Music Information Retrieval Conference}. 2019.

[8] Rafii, Zafar, and Bryan Pardo. ``Music/Voice Separation Using the Similarity Matrix.'' In {\it Proceedings of the 13th International Society for Music Information Retrieval Conference}. 2012.

[9] McFee, Brian, Colin Raffel, Dawen Liang, Daniel PW Ellis, Matt McVicar, Eric Battenberg, and Oriol Nieto. ``librosa: Audio and music signal analysis in python.'' In {\it Proceedings of the 14th python in science conference}, vol. 8. 2015.

[10] Panayotov, Vassil, Guoguo Chen, Daniel Povey, and Sanjeev Khudanpur. ``Librispeech: an ASR corpus based on public domain audio books.'' In {\it 2015 IEEE International Conference on Acoustics, Speech and Signal Processing (ICASSP)}, pp. 5206-5210. IEEE, 2015.

[11] Lenzo, Kevin. ``The CMU pronouncing dictionary.'' Carnegie Mellon University (2007).

[12] Holgate, Eric, Isabel Cachola, Daniel Preoţiuc-Pietro, and Junyi Jessy Li. ``Why swear? analyzing and inferring the intentions of vulgar expressions.'' In {\it Proceedings of the 2018 Conference on Empirical Methods in Natural Language Processing}, pp. 4405-4414. 2018.

\end{document}